\documentclass[]{aa}
\usepackage{amsmath,graphicx,amssymb}
\usepackage{txfonts}
\usepackage{natbib}
\usepackage{hyperref}

\newcommand{\Teff}{$T_\mathrm{eff}$}
\newcommand{\logg}{$\log g$}

\begin{document}

\title{Catalogue of stars measured in the Geneva seven-colour photometric system}

\author{E.~Paunzen\inst{1}}
\institute{Department of Theoretical Physics and Astrophysics, Masaryk University,
Kotl\'a\v{r}sk\'a 2, 611\,37 Brno, Czech Republic \\
\email{epaunzen@physics.muni.cz}
}

\date{}

\abstract
{The Geneva seven-colour photometric system is successfully applied to the study of various astrophysical 
objects. 
It measures the slope of the Paschen continuum, the Balmer discontinuity, and blocking absorption due to hydrogen or 
metallic lines.
One of its greatest strengths is its intrinsic homogeneity.}
{A new catalogue of the available measurements was generated, 30 years after the last publication.}
{The identifications for the individual stars were cross-checked on the basis of the $Gaia$ and 2MASS catalogues. 
The high precision coordinates together with proper motions (if available) are included, for the first time, in
the catalogue. Special caution was exercised with binaries and high-proper-motion stars.}
{The catalogue includes 42\,911 entries of highly accurate photometry.}
{The data of this catalogue can be used for various applications, such as new calibrations of astrophysical 
parameters, the
standardisation of new observations, and as additional information for ongoing and forthcoming all-sky surveys, 
such as the Transiting Exoplanet Survey Satellite (TESS).}

\keywords{Astronomical data bases -- Catalogs -- Stars: general -- Techniques: photometric}

\maketitle

\titlerunning{Catalogue -- Geneva 7-colour photometric system}
\authorrunning{Paunzen}

\section{Introduction} \label{introduction}

Photometry of astrophysical objects in different well-defined filters has been a powerful tool since
the introduction of photographic plates. With the invention of photomultipliers and charge-coupled devices,
new eras in astrophysics were triggered. Depending on the object type, photometric measurements
in different wavelength regions (not necessarily in the optical) 
can reveal significant characteristics with a low time investment. 
Unfortunately, the advantages of multi-filter usage are not used for satellite-based observations, such as
the Kepler \citep{2010ApJ...713L..79K} and Transiting Exoplanet Survey Satellite \citep[TESS;][]{2015JATIS...1a4003R} missions.

The homogeneity of a photometric system is most important when it comes to deriving calibrations of 
astrophysical parameters and model atmospheres, as well as isochrones \citep{2005ARA&A..43..293B}.  
A careful selection of standard stars and its replicability are essential for any standard system
\citep{2006MNRAS.373..781L,2021MNRAS.504.3730C}. 
The $Gaia$
consortium have taken a different approach to their photometric system. The definition of the three-filter
photometry of the Data Release 2 \citep[DR2;][]{2018A&A...616A...4E} is different than that of 
the Early Data Release 3 \citep[EDR3;][]{2021A&A...649A...3R} and will be changed again in the 
Data Release 3, which is unfortunate.

In this paper, on the other hand, we present observations within one of the most homogeneous photometric
system available, the Geneva seven-colour system \citep{1980VA.....24..141G}. Measurements were employed
under the leadership of the Observatoire de Gen{\`e}ve at six different 
observatories around the world. The filters were identical and the detectors well characterised. 
In addition, the reduction
process is well defined and all the data are processed in a homogeneous way. The disadvantage of
such a strategy is that the observed astrophysical objects are prescribed and limited.

The last published version of the Geneva seven-colour photometric catalogue dates back more than 30 years 
\citep{1989A&AS...78..469R}. Since then, regular updates have been made public via the 
General Catalogue of 
Photometric Data \citep[GCPD;][]{1997A&AS..124..349M}\footnote{\url{http://gcpd.physics.muni.cz/}}.

We present here the newest version (released September 2021), cross-matched with the recent $Gaia$
data releases and the Two Micron All Sky Survey (2MASS) in order to clearly identify the unique identifications
of the Geneva working group. This will allow, for the first time, the photometric data to be used in a much
more convenient and broader context.

\section{The Geneva seven-colour photometric system} \label{photometric_system}

We would like to give a short overview of the Geneva seven-colour photometric system. Detailed reviews
can be found in \citet{1972VA.....14...13G} and \citet{1980VA.....24..141G,1988A&A...206..357R}.

The system was defined in the late 1950s by Marcel Golay and first used at the Sphinx Observatory 
of the Jungfraujoch Scientific Station
in 1960. It has subsequently been applied at the Observatoire de Haute Provence,
the Gornergrat, Calar Alto, and La Silla observatories, and 
the IAC Observatory (Canary Islands). 

\begin{figure}
\begin{center}
\includegraphics[width=0.48\textwidth]{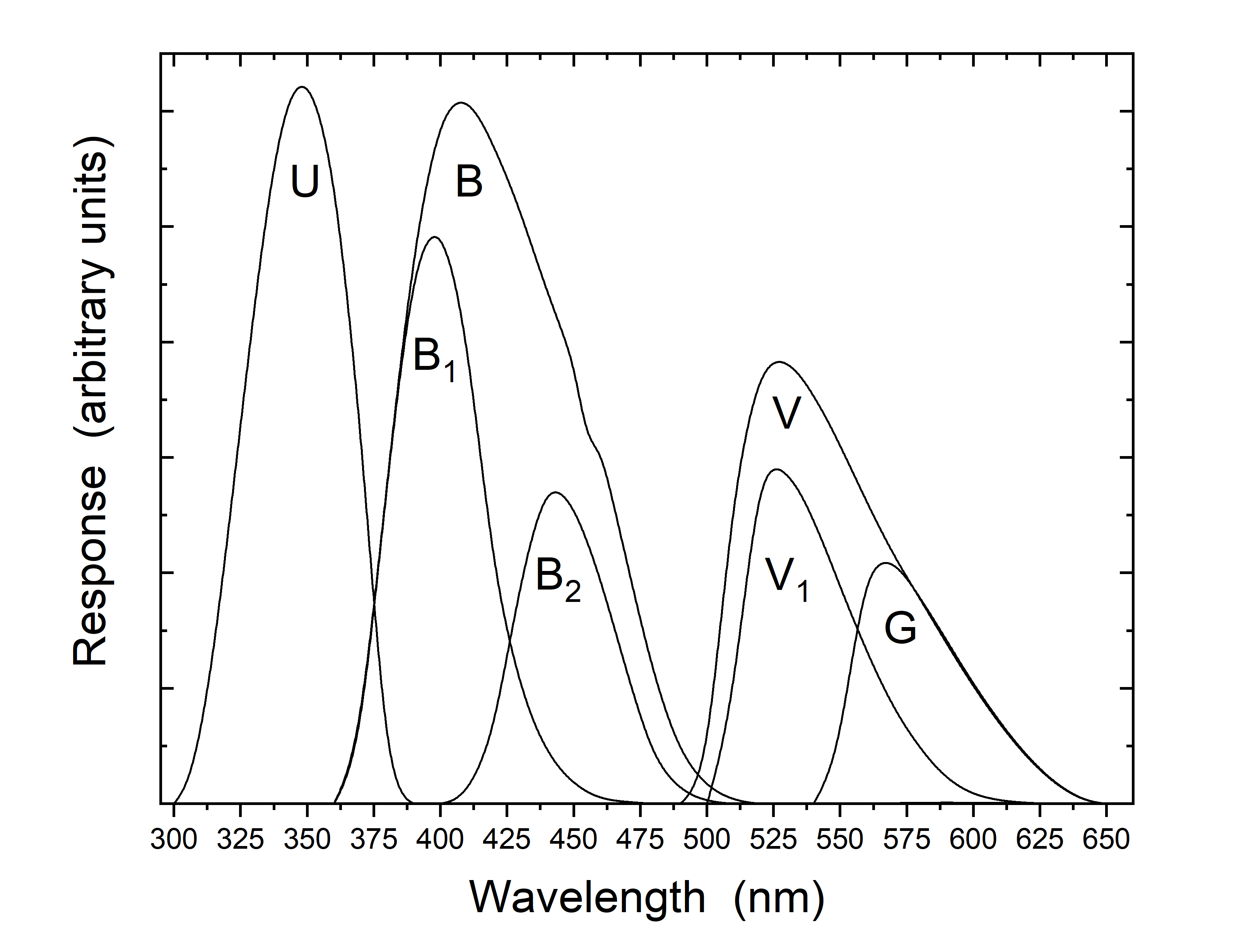}
\caption{Filter curves \citep[taken from][]{1988A&A...206..357R} of the seven filters.}
\label{passbands}
\end{center}
\end{figure}

The passbands (Fig. \ref{passbands}) were chosen with
the intent of reproducing the general properties of the, at that time, newly developed
and highly successful Johnson $UBV$ system. Whereas the $V$ filter is almost identical to the 
Johnson one, the $U$ and $B$ passbands are slightly adapted, presenting less overlap at the
Balmer jump of the hydrogen spectrum. The other four intermediate filters, $B_\mathrm{1}$, 
$B_\mathrm{2}$, $V_\mathrm{1}$,
and $G$, were added so that the stellar classification properties of spectrophotometry
involving the hydrogen absorption lines and gradients defined over the
Paschen continuum were well represented. Together they measure the wavelength region between 300 and
650\,nm.
The $B_\mathrm{1}$ and $B_\mathrm{2}$ filters each constitute a half part of the $B$ filter, and
likewise the $V_\mathrm{1}$ and $G$ filters are each a half part of the $V$ filter.

Several indices were defined, which are also included in the catalogue. Those indices are briefly described here.

First we have $U-B_\mathrm{2}$, $V_\mathrm{1}-G$, and $B_\mathrm{2}-V_\mathrm{1}$, which are the colour indices that
    are well correlated to the \Teff\ for different spectral types but are sensitive to the interstellar reddening.
    In addition, $U-B_\mathrm{2}$ and $B_\mathrm{2}-V_\mathrm{1}$ are almost insensitive to metallicity
    \citep{1980VA.....24..141G}, whereas $V_\mathrm{1}-G$ can be used to trace chemically peculiar stars
    \citep{1982A&A...114...23H}. 
    
Next we have $d$ and $\Delta$. Both are very similar to the $c_\mathrm{1}$ index of the Str{\"o}mgren-Crawford
    system \citep{1966ARA&A...4..433S} and the $Q$ parameter of the Johnson system \citep{1975PASP...87..805G}. 
    The $d$ index mainly measures the Balmer discontinuity and is 
    insensitive to interstellar extinction (as are $\Delta$, $g$, and $m_\mathrm{2}$). The loci of constant $d$ are 
    very similar to the loci of 
    constant $(U-B)$. The $\Delta$ index is sensitive to both chemical composition and gravity
    for cool-type stars.
    
    We also have $g$ and $m_\mathrm{2}$. These lines of constant values (isolines) are roughly parallel to 
    the \Teff\  axis for \Teff\,$>11\,000$\,K and
    are roughly parallel to the \logg\  axis for \Teff\,$<7\,000$\,K. Furthermore, both behave similarly  to 
    the $m_\mathrm{1}$ index of the Str{\"o}mgren-Crawford
    system \citep{1966ARA&A...4..433S}, that is, they are sensitive to the metallicity, measuring the 
    effect of line blocking in the region redder than 370\,nm. The difference between these two
    indices is the definition of how they were made insensitive to interstellar reddening.
    
Finally, we have $X$, $Y$, and $Z$. These reddening-free indices were introduced for stars hotter than 
    8500\,K through a rotation in the 3D space defined by the $d$, $\Delta$, and $g$ indices, 
    such that $X$ optimally correlates with \Teff\ for B-type stars and $Y$ correlates with \logg\ for the same
    spectral type range \citep{1979A&A....78..305C}. However, soon it was discovered that they are able to
    distinguish between cool-type dwarfs and giants \citep{1978prph.book.....G} as well as between normal and 
    chemically peculiar stars of the upper main sequence through the $Z$ index \citep{1980A&A....88..135C}, for example.

All types of `normal' stars were investigated within the Geneva photometric system, from the most massive ones 
\citep{1976A&A....48...87G,1993A&A...269..457C} to the lower main sequence \citep{1977PASP...89..706E} and giants 
\citep{1989Ap.....31..735K}. The success and capabilities are documented by the investigations of
binaries \citep{1998A&AS..132..367B}, Be stars \citep{1999A&A...346..134B},
chemically peculiar stars of the upper main sequence \citep{1980A&A....92..289H,1982A&A...114...23H}, 
Population II objects \citep{1986spcp.proc...53H}, 
shell stars \citep{1987A&A...177..193H,2000A&A...354..157H},
variables of all 
types \citep{1981A&A....97..274W,1992A&AS...95..471L,1994A&AS..108....9B,2000A&A...361..201E}, and
weak-lined F-type stars \citep{1991A&A...252..260H}, just to mention a few of the main star groups.

Several calibrations of the \Teff, \logg, absolute magnitude, and metallicity for the 
Geneva photometric system are available 
\citep{1997A&AS..122...51K,1999NewAR..43..343C,2005A&A...444..941P,2006A&A...458..293P,2017MNRAS.469.3042N}.

One interesting concept developed on the basis of the highly accurate measurements are the
`Geneva photometric boxes' \citep{1978A&A....62..189G,1980A&A....85..311C,1987A&A...177..233N}. 
With the availability of seven colours, a box around a so-called central star is defined as the 
set of stars that have similar colours or parameters in the vicinity of the central star. As a further step,
theoretical colours from stellar model atmospheres can also be used as a substitute for the central star. 
Then one can test the idea that two stars with almost the same colours (parameters) have almost the 
same physical properties. Here we illustrate this idea with an example. We consider stars within the classical
$\delta$ Scuti instability strip \citep{2016MNRAS.457.3163X} and their homogeneous photometric time 
series \citep{2019MNRAS.485.2380M}. We would like to determine how many stars are pulsating 
with a given amplitude if we take one box in this region as well as what can be used to discriminate between a 
pulsating star and an apparently constant star. With the help
of the presented catalogue, the new $Gaia$ data, and additional available photometric, as well as spectroscopic
measurements, such analyses can be done anew.

\begin{figure*}
\begin{center}
\includegraphics[width=0.9\textwidth]{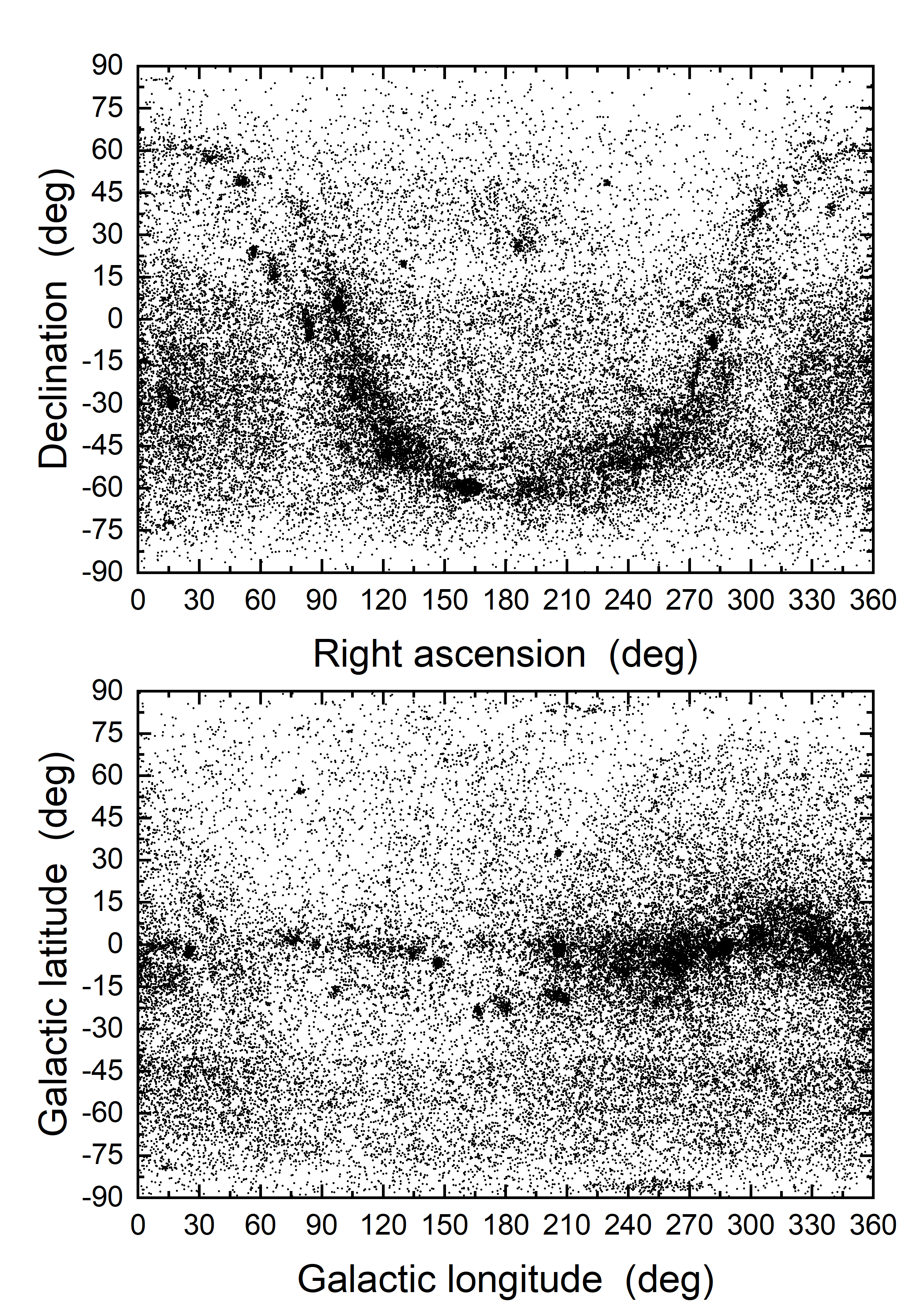}
\caption{Distribution of the catalogue stars in equatorial (upper panel) and Galactic (lower panel) coordinates.}
\label{coordinates}
\end{center}
\end{figure*}

\section{The catalogue} \label{catalogue}

The current version of the catalogue as available in the GCPD includes 43\,942 entries of standardised 
photometry. The last published one 
available in the SIMBAD Astronomical Database (II/169) consists of 29\,397 data points. The first important task was to unambiguously 
identify the individual entries on the basis of modern all-sky surveys (i.e. list the right ascension
and declination).

The unique numbering system of the GCPD, also used for the $uvby\beta$ catalogue by 
\citet{1998A&AS..129..431H}, for example, poses several problems and is somewhat anachronistic. 
Its designation `GEN\#' (as included in SIMBAD) is followed by a plus or minus sign, an integer, 
a dot, and then eight digits. It
is based on a hierarchical system of secondary catalogues and
numbering systems \citep{1989A&AS...78..469R}. There are special codes for members of
clusters, associations, different kinds of stars (for example, faint
blue stars, white dwarfs, and emission-line stars), and common
acronyms. To give just two examples, the high-proper-motion star Ross 413 (HIP 28940) is catalogued as
`GEN\# +9.80106025', whereas the Wolf-Rayet star WR 66 (HIP 74634) is listed under `GEN\# +6.10003891'.
There are three main problems with this identification system. The first is that the last version of the catalogue, published in 1988, includes only very inaccurate coordinates 
    in the Epoch 1900.0.\  Second, the secondary catalogues are sometimes not available, so no cross-match is possible.\ Finally, SIMBAD includes only a small fraction of the identifications, and sometimes they are incorrectly
    cross-matched. 

To overcome these severe limitations,
all objects were first cross-matched in the $Gaia$ DR2 \citep{2018A&A...616A..17A} 
and EDR3 \citep{2021A&A...649A...1G}. For this, we used the original identification and, if available, compared the Geneva
$V$ magnitude to the $Gaia$ $G$ one. 
For objects not found in the $Gaia$ data sets, we used the 2MASS catalogue \citep{2006AJ....131.1163S} as 
an additional source.

In total, 42\,911 entries could be unambiguously identified, all but 45 of which were in one the two $Gaia$ data releases. For those we list the EDR3 
(preferred) and DR2 identifications, the equatorial and Galactic coordinates, and the proper motions
(if available). For the remaining 45 stars, the 2MASS identifiers and coordinates are listed. With this information,
it is now possible to use the catalogue stars and cross-match them on the basis of precise coordinates. 

There are many high-proper-motion stars from the
compilations by \citet{1971lpms.book.....G} and \citet{1979lccs.book.....L} included in the catalogue that 
deserve special attention. It seems that the EDR3 has quite a lot of duplicate entries for single
objects. One example is G 270-159 (GEN\# +9.80270159), for which the EDR3 lists two sources within 0$\farcs$8:
2531243756196895104 ($G$\,=\,10.89\,mag) and 2531243756196895232 (14.70\,mag). The total proper motions
are 210.5 and 206.7 mas/yr, respectively. A visual inspection in different wavelength regions shows that
there is only one star of about 11th magnitude present. In such cases, the usage of the apparent magnitude 
allowed the duplicate entries to be sorted out. Nevertheless, the user is advised to check the entries in the
$Gaia$ catalogues when analysing high-proper-motion stars.

Stars that were found to be variable during the reduction process \citep{1982A&AS...48..503R} are marked with a `V'.

Another important point is the handling of binary systems. There are visual binary systems that normally
have different entries -- `A' and `B', for example -- and spectroscopic binary systems with an entry `AB'.
If possible, separated systems have been assigned the correct identification. A good example is the
high-proper-motion visual binary 35 Psc (GEN\# +1.00001061), which has three entries: GEN\# +1.00001061 AV
(and not GEN\# +1.00001061A as listed in SIMBAD), GEN\# +1.00001061 B, and GEN\# +1.00001061 ABV. 
They were assigned to Gaia EDR3 2752338227234710784 (AV and ABV) Gaia EDR3 2752338227234710912 (B). In general,
the users are advised to check the individual entries for known binary systems (second column in the catalogue).

\begin{figure}
\begin{center}
\includegraphics[width=0.48\textwidth]{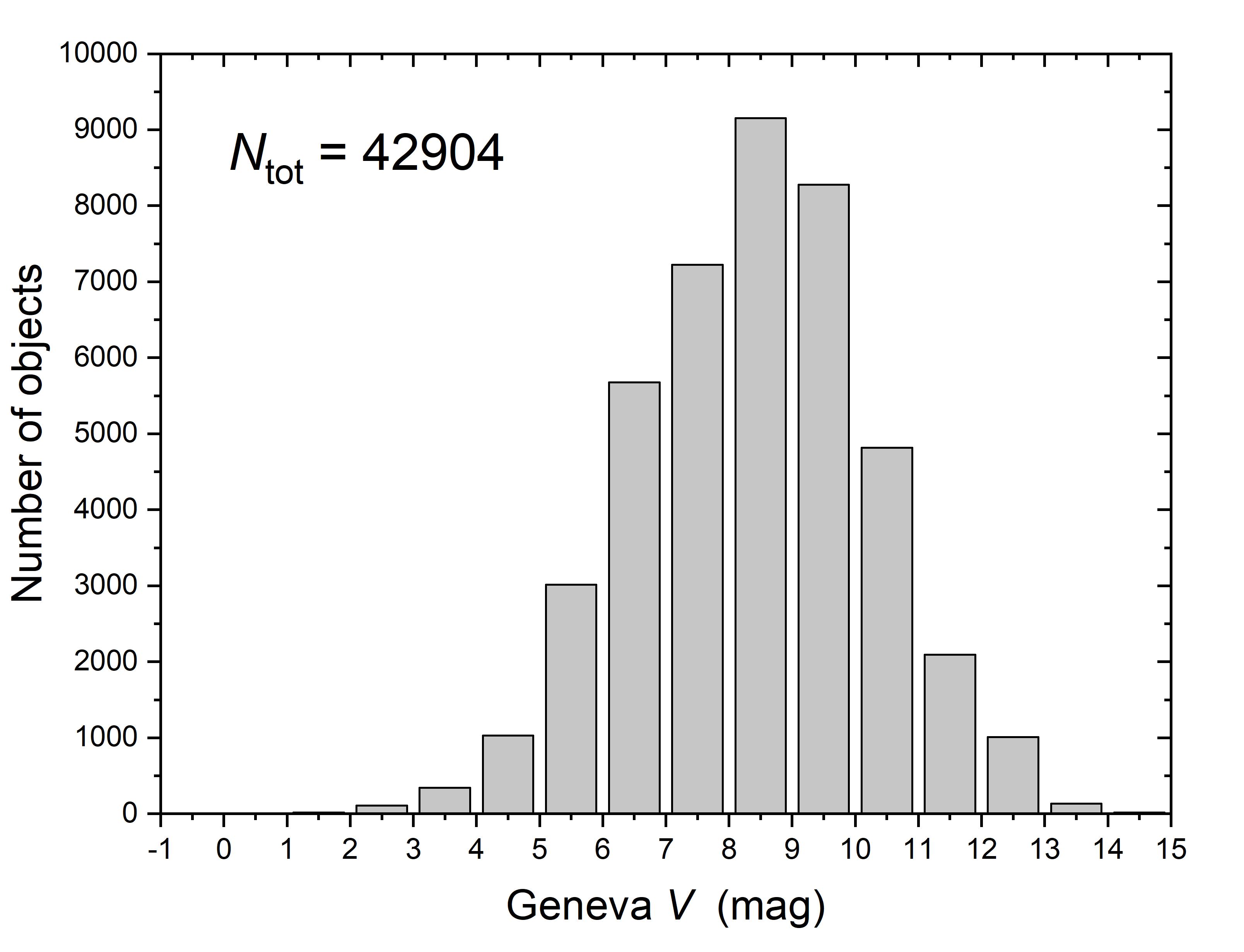}
\caption{Geneva $V$ distribution of 42\,904 catalogue entries. Seven do not have one.}
\label{histogram_V}
\end{center}
\end{figure}

The final catalogue includes 42\,911 entries and will be provided in electronic form in
the VizieR (CDS) database and the GCPD. It is organised as follows:

\begin{itemize}
\item Column 1: Identification in the GCPD.
\item Column 2: Remarks on duplicity, variability, and so on.
\item Column 3: Geneva $V$ magnitude.
\item Column 4: $U-B$.
\item Column 5: $V-B$.
\item Column 6: $B_\mathrm{1}-B$.
\item Column 7: $B_\mathrm{2}-B$.
\item Column 8: $V_\mathrm{1}-B$.
\item Column 9: $G-B$.
\item Column 10: $U-B_\mathrm{1}$.
\item Column 11: $U-B_\mathrm{2}$.
\item Column 12: $B_\mathrm{1}-B_\mathrm{2}$.
\item Column 13: $V_\mathrm{1}-G$.
\item Column 14: $B_\mathrm{2}-G$.
\item Column 15: $B_\mathrm{2}-V_\mathrm{1}$.
\item Column 16: $d = (U-B_\mathrm{1}) - 1.430(B_\mathrm{1}-B_\mathrm{2})$.
\item Column 17: $\Delta = (U-B_\mathrm{2}) - 0.832(B_\mathrm{2}-G)$.
\item Column 18: $g = (B_\mathrm{1}-B_\mathrm{2}) - 1.357(V_\mathrm{1}-G)$.
\item Column 19: $m_\mathrm{2} = (B_\mathrm{1}-B_\mathrm{2}) - 0.457(B_\mathrm{2}-V_\mathrm{1})$.
\item Column 20: $X = +0.3788 + 1.3764U - 1.2162B_\mathrm{1} - 0.8498B_\mathrm{2} - 0.1554V_\mathrm{1} + 0.8450G$.
\item Column 21: $Y = -0.8288 + 0.3235U - 2.3228B_\mathrm{1} + 2.3363B_\mathrm{2} + 0.7495V_\mathrm{1} - 1.0865G$.
\item Column 22: $Z = -0.4572 + 0.0255U - 0.1740B_\mathrm{1} + 0.4696B_\mathrm{2} - 1.1205V_\mathrm{1} + 0.7994G$.
\item Column 23: Weight of Geneva $V$ magnitude.
\item Column 24: Standard deviation of Geneva $V$ magnitude.
\item Column 25: Weight of the colours.
\item Column 26: Mean standard deviation for the colours.
\item Column 27: $Gaia$ or 2MASS identification.
\item Column 28: Right ascension (J2000).
\item Column 29: Declination (J2000).
\item Column 30: Proper motion in right ascension direction.
\item Column 31: Standard error of proper motion in right ascension direction.
\item Column 32: Proper motion in declination direction.
\item Column 33: Standard error of proper motion in declination direction.
\item Column 34: Galactic longitude.
\item Column 35: Galactic latitude.
\end{itemize}

Because of the large number of columns, we do not give an example here. 

\section{Sample characteristics and conclusions} \label{conclusions}

Figure \ref{coordinates} shows the distribution of the objects on the sky in equatorial and Galactic coordinates.
The Galactic disk and several open clusters are immediately visible. There is also a high concentration of
stars with Galactic latitudes $< -40\degr$ (i.e. in the direction of the South Galactic Pole). There is a
significant number of high-proper-motion stars among these objects. With the $Gaia$ data, these
objects are interesting targets for a new analysis.

The distribution of the Geneva $V$ magnitudes of all but seven catalogue entries
is shown in Fig. \ref{histogram_V}. Of the stars listed in the catalogue, 98\% are in the magnitude range of 4.0\,$<$\,$V$\,$<$\,13.0\,mag, 
with a peak at about 8.5\,mag.
This is exactly the magnitude range in which the TESS satellite, for example, performs high precision
photometric time series observations. The Geneva seven-colour photometry can serve as an excellent
supplement to these observations for determining astrophysical parameters. 

\begin{figure}
\begin{center}
\includegraphics[width=0.48\textwidth]{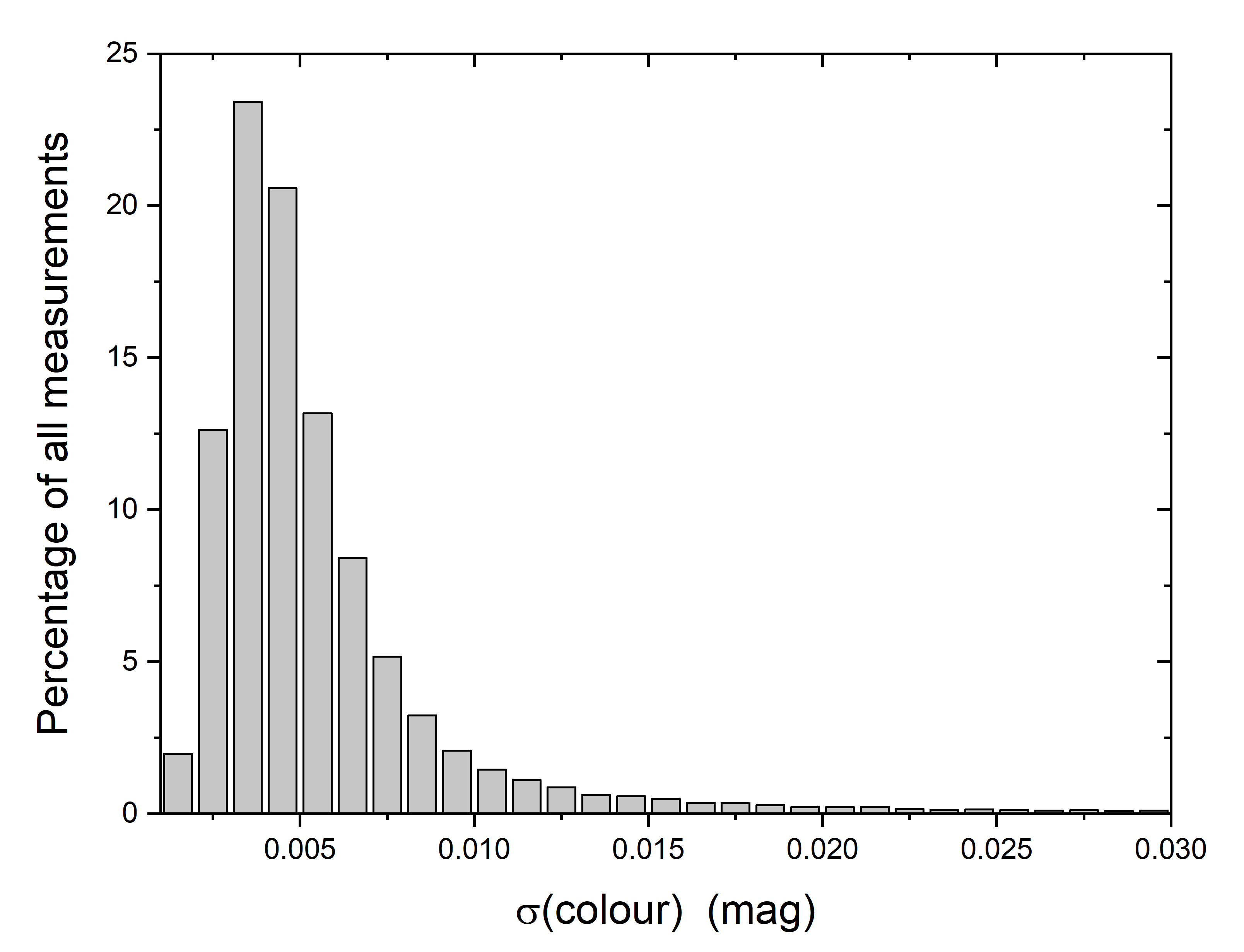}
\caption{Distribution of the mean standard deviation for the colours. About 92\% of all 
measurements have a standard deviation below 0.01\,mag.}
\label{sigma_colour}
\end{center}
\end{figure}

The high precision of the available photometric data is documented in Fig. \ref{sigma_colour}.
From the complete sample, about 92\% of all measurements have a standard deviation below 0.01\,mag, and
80\% even have a standard deviation below 0.005\,mag. This allows one to establish very accurate calibrations.
Furthermore, the data can be used for fitting the       spectral energy distribution (SED), for example within 
the VO SED Analyzer 
(VOSA) tool \citep{2008A&A...492..277B}. 

\begin{figure*}
\begin{center}
\includegraphics[width=0.97\textwidth]{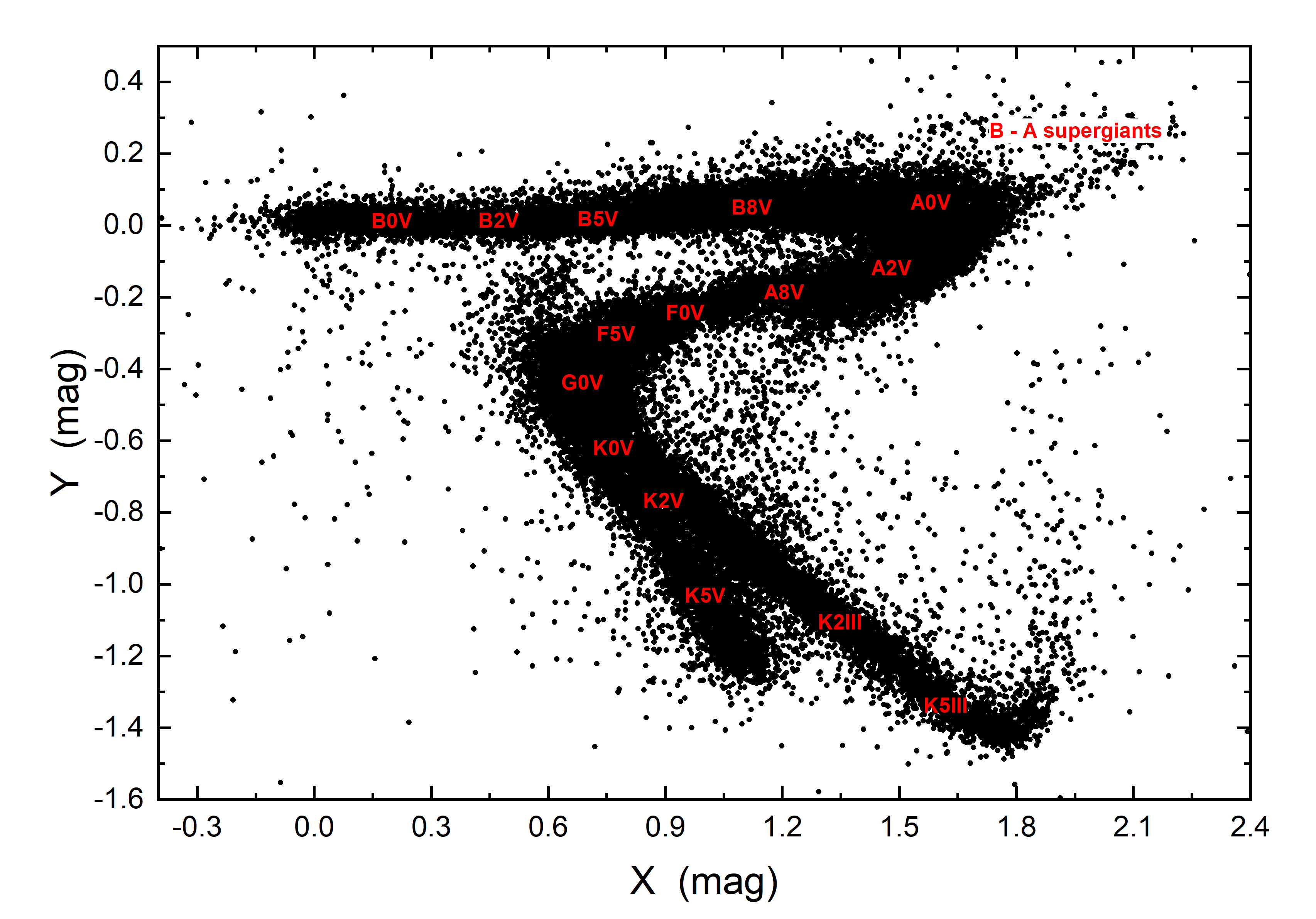}
\caption{The $X$ versus $Y$ diagram of the catalogue stars. The spectral types 
\citep[taken from][]{1994ASPC...60..157H} are listed for guidance.}
\label{XY}
\end{center}
\end{figure*}

Finally, Fig. \ref{XY} shows the diagram of reddening-free $X$ and $Y$ indices as described in 
Sect. \ref{photometric_system} for the complete sample. As guidance, the spectral types taken
from \citet{1994ASPC...60..157H} are also included. The broad main sequence band and the bifurcation of cool-type
giants are clearly visible. The complete spectral range is very well represented. 
This diagram, together with several others as described in more detail in
\citet{1980VA.....24..141G}, allows one to establish detailed calibrations of astrophysical stellar parameters. 

The data of this catalogue can be used for new calibrations of astrophysical
parameters as well as for the standardisation of new
observations.\ They can also be used as additional information for ongoing as well
as forthcoming all-sky surveys.

\begin{acknowledgements}
This work has made use of data from the European Space Agency (ESA) mission 
{\it Gaia} (\url{https://www.cosmos.esa.int/gaia}), processed by the {\it Gaia} Data 
Processing and Analysis Consortium (DPAC, \url{https://www.cosmos.esa.int/web/gaia/dpac/consortium}). 
Funding for the DPAC has been provided by national institutions, in particular the institutions
participating in the {\it Gaia} Multilateral Agreement. This research has made use of the SIMBAD database,
operated at CDS, Strasbourg, France and 
of the WEBDA data base, operated at the Department of Theoretical
Physics and Astrophysics of the Masaryk University.
\end{acknowledgements}

\bibliographystyle{aa}
\bibliography{Geneva_catalog}

\end{document}